\documentclass[journal]{IEEEtran}
\usepackage{graphicx}
\usepackage{amssymb}
\usepackage{multirow}
\usepackage{tabularx}
\usepackage{acronym}
\usepackage{color}
\usepackage{lettrine}
\usepackage{mdwlist}
\usepackage{gensymb}
\usepackage{paralist}
\usepackage{cite}
\usepackage[cmex10]{amsmath}
\usepackage[font=footnotesize]{caption}
\usepackage[font=footnotesize]{subcaption}
\usepackage[utf8]{inputenc}
\usepackage{url}
\usepackage[utf8]{inputenc}
\usepackage[english]{babel}
\usepackage{amsthm}
\usepackage{hyperref}
\usepackage[ruled,vlined]{algorithm2e}

\usepackage[T1]{fontenc}

\ifCLASSINFOpdf
\else
\fi
\usepackage{amsmath}
\usepackage[cmintegrals]{newtxmath}
\begin{document}
	%
	\title{100~Gbps Indoor Access and 4.8~Gbps Outdoor Point-to-Point LiFi Transmission Systems using Laser-based Light Sources}
	%
	%
	%
	
	\author{Cheng~Cheng, Sovan~Das, Stefan~Videv, Adrian~Spark, Sina~Babadi, Aravindh~Krishnamoorthy,~\IEEEmembership{Graduate Student Member,~IEEE}, Changmin~Lee, Daniel~Grieder, Kathleen~Hartnett, Paul~Rudy, James~Raring, Marzieh~Najafi,~\IEEEmembership{Student Member,~IEEE}, Vasilis~K.~Papanikolaou,~\IEEEmembership{Member,~IEEE}, Robert~Schober,~\IEEEmembership{Fellow,~IEEE}~and~Harald~Haas,~\IEEEmembership{Fellow,~IEEE}
		\thanks{Cheng~Cheng, Stefan Videv, Adrian Spark, Sina Babadi~and~Harald Haas are with LiFi R\&D centre ,the Department
			of Electronic \& Electrical Engineering, University of Strathclyde, Glasgow,
			UK, G1 1XQ}
		\thanks{Changmin Lee, Sovan Das, Daniel Grieder, Kathleen Hartnett, Paul Rudy and James Raring are with Kyocera SLD Laser, Inc. 111 Castilian Dr, Goleta, California, USA, 93117}
		\thanks{Aravindh Krishnamoorthy, Marzieh Najafi, Vasilis K. Papanikolaou and Robert Schober are with Institute for Digital Communications, Friedrich-Alexander-Universit\"{a}t Erlangen-N\"{u}rnberg, Germany, 91054}
	\thanks{Cheng Chen, Stefan Videv, Adrian Sparks, Sina Babadi and Harald Haas acknowledge partial financial support from Kyocera SLD Laser, the Department of Science, Innovation \& Technology in the UK through the project: Enabling Architectures and Solutions for Open Networks (REASON). Harald Haas additionally acknowledges the Alexander von Humboldt Foundation in Germany for a research award.}
}

	%
	%

	\markboth{Journal of Lightwave Technology}%
	{Shell \MakeLowercase{\textit{et al.}}: Bare Demo of IEEEtran.cls for IEEE Journals}
	%



\makeatletter
\newif\ifAC@uppercase@first%
\def\Aclp#1{\AC@uppercase@firsttrue\aclp{#1}\AC@uppercase@firstfalse}%
\def\AC@aclp#1{%
	\ifcsname fn@#1@PL\endcsname%
	\ifAC@uppercase@first%
	\expandafter\expandafter\expandafter\MakeUppercase\csname fn@#1@PL\endcsname%
	\else%
	\csname fn@#1@PL\endcsname%
	\fi%
	\else%
	\AC@acl{#1}s%
	\fi%
}%
\def\Acp#1{\AC@uppercase@firsttrue\acp{#1}\AC@uppercase@firstfalse}%
\def\AC@acp#1{%
	\ifcsname fn@#1@PL\endcsname%
	\ifAC@uppercase@first%
	\expandafter\expandafter\expandafter\MakeUppercase\csname fn@#1@PL\endcsname%
	\else%
	\csname fn@#1@PL\endcsname%
	\fi%
	\else%
	\AC@ac{#1}s%
	\fi%
}%
\def\Acfp#1{\AC@uppercase@firsttrue\acfp{#1}\AC@uppercase@firstfalse}%
\def\AC@acfp#1{%
	\ifcsname fn@#1@PL\endcsname%
	\ifAC@uppercase@first%
	\expandafter\expandafter\expandafter\MakeUppercase\csname fn@#1@PL\endcsname%
	\else%
	\csname fn@#1@PL\endcsname%
	\fi%
	\else%
	\AC@acf{#1}s%
	\fi%
}%
\def\Acsp#1{\AC@uppercase@firsttrue\acsp{#1}\AC@uppercase@firstfalse}%
\def\AC@acsp#1{%
	\ifcsname fn@#1@PL\endcsname%
	\ifAC@uppercase@first%
	\expandafter\expandafter\expandafter\MakeUppercase\csname fn@#1@PL\endcsname%
	\else%
	\csname fn@#1@PL\endcsname%
	\fi%
	\else%
	\AC@acs{#1}s%
	\fi%
}%
\edef\AC@uppercase@write{\string\ifAC@uppercase@first\string\expandafter\string\MakeUppercase\string\fi\space}%
\def\AC@acrodef#1[#2]#3{%
	\@bsphack%
	\protected@write\@auxout{}{%
		\string\newacro{#1}[#2]{\AC@uppercase@write #3}%
	}\@esphack%
}%
\def\Acl#1{\AC@uppercase@firsttrue\acl{#1}\AC@uppercase@firstfalse}
\def\Acf#1{\AC@uppercase@firsttrue\acf{#1}\AC@uppercase@firstfalse}
\def\Ac#1{\AC@uppercase@firsttrue\ac{#1}\AC@uppercase@firstfalse}
\def\Acs#1{\AC@uppercase@firsttrue\acs{#1}\AC@uppercase@firstfalse}
\robustify\Ac
\robustify\Aclp
\robustify\Acfp
\robustify\Acp
\robustify\Acsp
\robustify\Acl
\robustify\Acf
\robustify\Acs
\def\AC@@acro#1[#2]#3{%
	\ifAC@nolist%
	\else%
	\ifAC@printonlyused%
	\expandafter\ifx\csname acused@#1\endcsname\AC@used%
	\item[\protect\AC@hypertarget{#1}{\acsfont{#2}}] #3%
	\ifAC@withpage%
	\expandafter\ifx\csname r@acro:#1\endcsname\relax%
	\PackageInfo{acronym}{%
		Acronym #1 used in text but not spelled out in
		full in text}%
	\else%
	\dotfill\pageref{acro:#1}%
	\fi\\%
	\fi%
	\fi%
	\else%
	\item[\protect\AC@hypertarget{#1}{\acsfont{#2}}] #3%
	\fi%
	\fi%
	\begingroup
	\def\acroextra##1{}%
	\@bsphack
	\protected@write\@auxout{}%
	{\string\newacro{#1}[\string\AC@hyperlink{#1}{#2}]{\AC@uppercase@write #3}}%
	\@esphack
	\endgroup}
\makeatother	

\acrodef{6g}[6G]{6th generation}
\acrodef{lifi}[LiFi]{light fidelity}
\acrodef{awg}[AWG]{arbitary waveform generator}
\acrodef{smd}[SMD]{surface mount device}
\acrodef{dc}[DC]{direct current}
\acrodef{ac}[AC]{alternating current}	
\acrodef{ld}[LD]{laser diode}
\acrodef{pin}[PIN]{positive-intrinsic-negative}	
\acrodef{pc}[PC]{personal computer}	
\acrodef{ofdm}[OFDM]{orthogonal frequency division multiplexing}
\acrodef{qam}[QAM]{quadrature amplitude modulation}	
\acrodef{dco}[DCO]{DC-biased optical}
\acrodef{fft}[FFT]{fast Fourier transform}
\acrodef{ifft}[IFFT]{inverse fast Fourier transform}
\acrodef{rrc}[RRC]{root-square raised cosine}
\acrodef{papr}[PAPR]{peak-to-average power ratio}
\acrodef{cp}[CP]{cyclic prefix}
\acrodef{rls}[RLS]{recursive least square}
\acrodef{ber}[BER]{bit error rate}
\acrodef{hh}[HH]{Hughes-Hartogs}
\acrodef{snr}[SNR]{signal-to-noise ratio}
\acrodef{lte}[LTE]{Long-Term Evolution}
\acrodef{pd}[PD]{photodiode}
\acrodef{ir}[IR]{infrared}
\acrodef{wdm}[WDM]{wavelength division multiplexing}
\acrodef{vr}[VR]{virtual reality}
\acrodef{ar}[AR]{augmented reality}
\acrodef{rf}[RF]{radio frequency}
\acrodef{iov}[IoV]{Internet-of-Vehicle}
\acrodef{its}[ITS]{Intelligent transportation system}
\acrodef{led}[LED]{light-emitting diode}
\acrodef{lidar}[LiDAR]{light detection and ranging}
\acrodef{gan}[GaN]{Gallium nitride}
\acrodef{ssl}[SSL]{solid state lighting}
\acrodef{mmwave}[mmWave]{millimeter-wave}
\acrodef{sld}[SLD]{superluminescent diode}
\acrodef{ook}[OOK]{on-off keying}
\acrodef{apd}[APD]{avalanche photodiode}
\acrodef{los}[LOS]{line-of-sight}
\acrodef{na}[NA]{numerical aperture}
\acrodef{fec}[FEC]{forward error correction}
\acrodef{rgb}[RGB]{red-green-blue}
\acrodef{cad}[CAD]{computer-aided design}

\maketitle

\begin{abstract}
In this paper, we demonstrate the communication capabilities of light-fidelity (LiFi) systems based on high-brightness and high-bandwidth integrated laser-based sources in a surface mount device (SMD) packaging platform. The laser-based source is able to deliver 450 lumens of white light illumination and the resultant light brightness is over $\bf 1000~cd/mm^2$. It is demonstrated that a wavelength division multiplexing (WDM) LiFi system with ten parallel channels is able to deliver over 100~Gbps data rate with the assistance of Volterra filter-based nonlinear equalisers. In addition, an aggregated transmission data rate of 4.8~Gbps has been achieved over a link distance of 500~m with the same type of SMD light source. This work demonstrates the scalability of LiFi systems that employ laser-based light sources, particularly in their capacity to enable high-speed short range, as well as long-range data transmission.
\end{abstract}


\section{Introduction}
\IEEEPARstart{L}{asers} have been widely adopted and have been consistently used to develop new technologies in many fields, such as telecommunications, remote sensing and chemistry \cite{barnoski2012fundamentals,9455394}. 
In recent decades, visible light \acp{ld} using \ac{gan} material have been developed and have enabled several useful applications such as blue ray disc and high quality projection display systems \cite{Ichimura_2000,Chen:17}. Recently, \ac{gan}-based \acp{ld} have been developed for \ac{ssl} applications \cite{ma10101166}. With the excellent energy-efficiency, \ac{gan}-based \ac{led} lamps and light bulbs have been widely commercialised to replace the old incandescent lighting infrastructure. Compared to \ac{gan}-based \ac{led} illumination capabilities, \ac{gan}-based \acp{ld} can offer more than ten times higher brightness, longer lighting range, improved directionality and compact device dimension, which makes them suitable for scenarios where it is challenging for \ac{gan}-based \acp{led} to achieve the desired performance. Several potential applications of \ac{gan}-based \acp{ld} include \ac{vr}/\ac{ar} display, automotive lighting and advanced medical devices \cite{tsao2014toward,blue_laser_applications}.

In recent years, the concept of using light for wireless networking, \ac{lifi}, has been extensively investigated in both academia and industry \cite{Haas2016}. A major advantage of this technology is that the infrastructure can provide both illumination and wireless communication functionalities concurrently. Compared to conventional \ac{rf} wireless communication technologies, \ac{lifi} systems use licence-free spectrum to transmit signals. In addition, \ac{lifi} can handle dense scenarios and can deliver confidential information securely. Compared to other new radio technologies, such as \ac{mmwave} and Sub-THz systems, \ac{lifi} is advantageous in terms of system complexity. For example, the directional transmission in \ac{mmwave} or Sub-THz systems is enabled by beamforming techniques which require complex signal processing and multiple antennas/\ac{rf} chains. In contrast, \ac{lifi} transmitters are inherently directional and so they require no additional hardware and signal processing. Therefore, \ac{lifi} has been considered to be a promising candidate in the \ac{6g} networks \cite{9170653}. Most of the existing \ac{lifi} studies are based on \acp{led}, which pose limited modulation bandwidths in the range of a few to tens of MHz \cite{7072557}. This significantly limits the achievable rate of \ac{lifi} systems. With research efforts focusing on how to extend the bandwidth of light sources, \ac{gan}-based micro \acp{led} and \acp{sld} have been developed to deliver wireless data rates of over 10~Gbps \cite{9040548,Shen:16}. Despite the transmission rate improvement, micro \acp{led} can emit a very low optical power, which makes it difficult to provide effective illumination and long range transmission. Therefore, laser-based light sources have been proposed for \ac{lifi} transmission \cite{tsonev2015towards}. \Acp{ld} can provide dramatically improved brightness and a higher modulation bandwidth of more than 1~GHz compared to \acp{led}. These features make it possible to develop \ac{lifi} systems that can provide both high-brightness illumination and ultra-high speed/long range data transmission.

There are several studies in the literature reporting results of \ac{lifi} research based on laser sources. \Ac{gan}-based \acp{ld} with phosphor have been considered to achieve wireless transmissions with high data rates of a few Gbps using \ac{ook} modulation \cite{Lee:15,lee2016dynamic} and \ac{ofdm} \cite{chi2015phosphorous,7247373}. A recent study considered using a free-form multipath lens to make laser-based \ac{lifi} transmission fulfil the eye-safety constraint \cite{9743382}, which is able to deliver a transmission  data rate of over 1~Gbps and a high optical power output at the same time. The use of \ac{rgb} lasers to establish white light illumination has been considered in \cite{Janjua:15}, which also utilises the differences in wavelength to form \ac{wdm} transmission so that the aggregated data rate can be improved. Our previous work has demonstrated the use of a dual \ac{ld} unit with a \ac{smd} package to develop a two-channel \ac{wdm} \ac{lifi} system achieving a data rate of 26~Gbps with \ac{ofdm} and high brightness white light illumination \cite{9601315}. 
Recently, we have reported a ten channel \ac{wdm} \ac{lifi} system with the improved version of the \ac{smd} laser sources in \cite{ECOC2023}. In this paper, we extend the work in \cite{ECOC2023} by including additional technical results, insights and discussions. 
By combining the data rates achieved by \ac{smd} laser sources with ten wavelengths (including blue and infrared), an aggregate data rate of over 100~Gbps can be achieved. In addition, with the overlapped coverage of multiple \ac{smd} laser sources, the overall illumination range and brightness can be enhanced. Furthermore, we showcase a \ac{lifi} transmission based on the \ac{smd} laser source in an outdoor environment with a link distance of 500~m, which achieved an aggregated data rate of 4.84~Gbps. 

The remainder of this paper is structured as follows: The potential use cases of the demonstrated \ac{lifi} system is discussed in Section~\ref{sec:use_cases}. The schematic and design of the dual \acp{ld} \ac{smd} package is introduced in Section~\ref{sec:smd_laser}. The \ac{lifi} system based on \ac{ofdm} and nonlinear equaliser with a single channel is presented in Section~\ref{sec:single_LiFi}. The 100~Gbps \ac{lifi} system with ten \ac{wdm} channels are presented in Section~\ref{sec:wdm_LiFi}. The multi-Gbps data rate 500~m \ac{lifi} system is presented in Section~\ref{sec:500m_LiFi}. Finally, the conclusion is drawn in Section~\ref{sec:conclusion}.

\begin{figure}[!t]
	\begin{center}
		\includegraphics[width=0.45\textwidth]{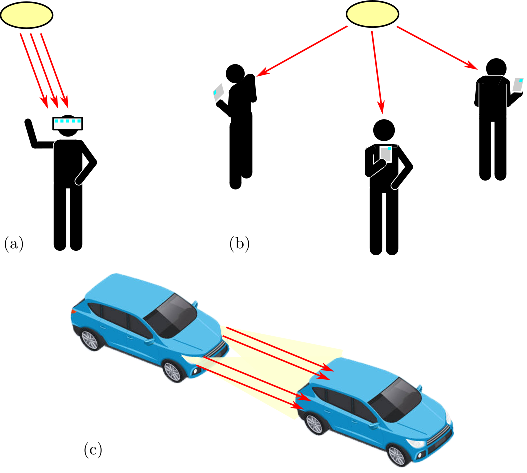}
	\end{center}
	\caption{High-speed and long range \ac{lifi} use cases: (a) Indoor point-to-point links (b) Indoor point to multi-points links (c) Outdoor vehicle-to-vehicle links}
	\label{fig:use_cases}
\end{figure}

\begin{figure}[!t]
	\begin{center}
		\includegraphics[width=0.5\textwidth]{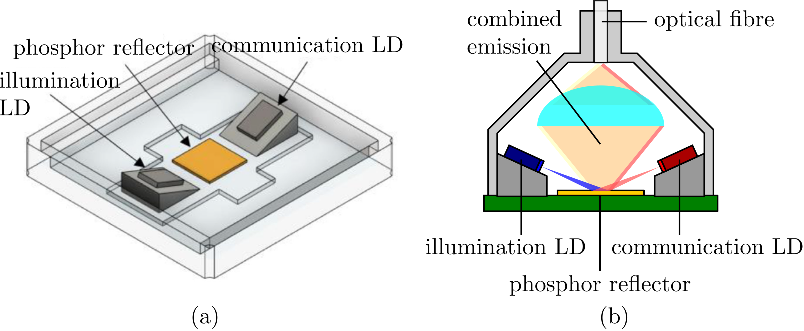}
	\end{center}
	\caption{(a) Schematic of the \ac{smd} light source with two LDs operating for illumination and communication functionalities. (b) Cross-section schematic of the \ac{smd} package.}
	\label{fig:SMD_package}
\end{figure}
\section{LiFi use cases with High-speed  parallel transmission and long range transmission}
\label{sec:use_cases}
In this section, The potential use cases of the considered \ac{lifi} systems are discussed. In the indoor scenarios, there are many wireless link use cases where ultra-high data rate is required. For example, the future \ac{vr} will offer a life-like immersive experience, where the image resolution will be in the range of 8K to 12K \cite{cuervo2018creating}. With the signal processing moving to the mobile edge, the required data rate will increase to tens of/hundreds of Gbps. In this case, the point-to-point high-speed \ac{lifi} parallel transmission can be used to fulfil the requirement, as shown in Figure~\ref{fig:use_cases}~(a). Another useful point-to-point use case is where large files need to be transferred between two mobile devices. The high speed \ac{lifi} parallel transmission is able to significantly shorten the transmission time to a few seconds. Additionally, the transmission is secure and wireless. Providing high-speed wireless broadband to a large number of mobile users is an important use case for point-to-multi-point scenarios, as shown in Figure~\ref{fig:use_cases}~(b). This is motivated by the extremely short reuse distance of \ac{lifi} as the light sources are directional and causes very little interferences to nearby users \cite{cbh1601}. In conjunction with the wide bandwidth of the laser source, a very high data density can be achieved in this use case. High speed wireless point-to-multi-point links in Industrial 4.0 is another use case for the considered \ac{lifi} system, especially in factories where \ac{rf} transmission is highly restricted. The multi-Gbps long range transmission capability of the laser-based \ac{lifi} will be useful in the outdoor vehicle-to-vehicle data transmission, as shown in Figure~\ref{fig:use_cases}~(c). In recent years, laser-based headlights have been developed for automotive applications, which provide a great platform for deploying the demonstrated \ac{lifi} system to establish high-speed data transmission between vehicles for future \ac{iov} and \ac{its} \cite{7932857}. Last mile wireless backhaul links for wireless networks and wireless fronthaul links for distributed antenna systems are potential use cases of long range \ac{lifi} systems, as well.

\begin{figure*}[!t]
	\begin{center}
		\includegraphics[width=1\textwidth]{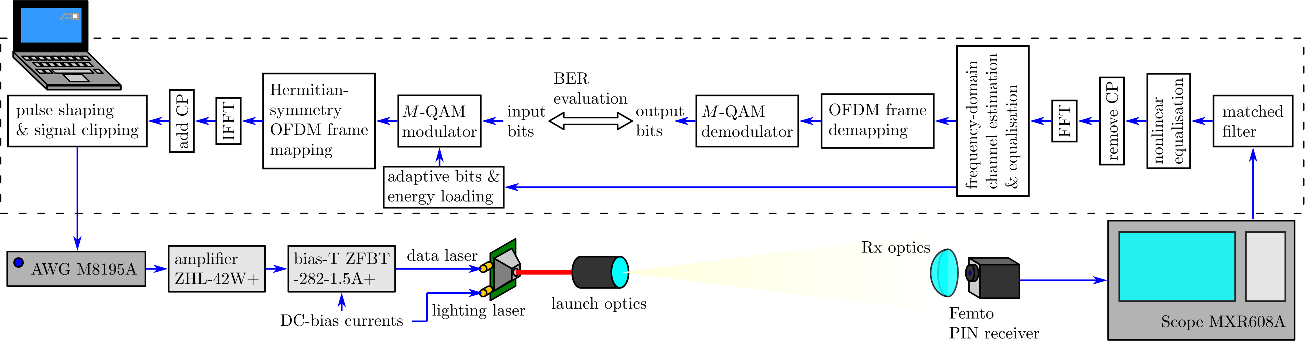}
	\end{center}
	\caption{Block diagram of a single link \ac{lifi} system}
	\label{fig:single_setup}
\end{figure*}

\section{Dual wavelength laser integrated surface mounting device}
\label{sec:smd_laser}
In this section, the design of the \ac{smd} laser source is introduced. The structure of the device package is similar to that presented in \cite{9601315}. Two \acp{ld} are deployed on the two wedges mounted on the edges of the package and a phosphor reflector is deployed in the centre of the package, as shown in Figure~\ref{fig:SMD_package}~(a). The orientations of the \acp{ld} are configured so that the emitted collimated light incident to the phosphor reflector and the reflected light is coupled into an optical fibre, as shown in Figure~\ref{fig:SMD_package}~(b). One of the \acp{ld} emitting blue light (450~nm) is designed for illumination purposes. When the blue light incident to the phosphor reflector, part of the blue light is diffusely reflected, and part of the light excites the phosphor material to emit yellow light. The combination of blue and yellow light forms the white emission for illumination, which uses the same principle of the white \ac{led} for lighting \cite{7072557}. Note that the phosphor reflector not only diffuses the collimated light from \acp{ld} but also converts the coherent light to non-coherent light, which is important for eye-safety considerations and illumination performance. The second \ac{ld} emitting light of a different wavelength is designed for communication purposes. In order to establish a system with multiple \ac{wdm} channels, \acp{ld} from different materials are used so that different copies of the \acp{smd} can emit light of different wavelengths for communication. In this work, ten devices with ten wavelengths have been made which include three blue \acp{ld} (405~nm, 450~nm, 455~nm) and seven infrared \acp{ld} (850~nm, 900~nm, 905~nm, 940~nm, 955~nm, 980~nm, 1064~nm). The light beams from illumination and communication \acp{ld} form an overlapped emission from the phosphor reflector. 
Compared to conventional \ac{led}-based light sources, the presented \ac{smd} laser source offers much higher optical power and a dramatic increase in modulation bandwidth for \ac{lifi}.


\section{High speed LiFi transmission with a single channel}
\label{sec:single_LiFi}
The multi-channel \ac{lifi} transmission is composed of a number of single links with various wavelengths in parallel, which share a similar system deign. In this section, the details of experimental design and the performance of a single \ac{lifi} link are presented. A block diagram of a single link \ac{lifi} system is shown in Figure~\ref{fig:single_setup}. The optical link system is composed of software and hardware parts. The transmitter hardware includes a high-speed \ac{awg}, Keysight M8195A, to convert digital signals to high-speed analogue waveform, a power amplifier, Mini-Circuit ZHL-42W+, to boost the modulated \ac{ac} signal, a bias-T, Mini-Circuit ZFBT-282-1.5A+, to combine the \ac{dc}-bias and \ac{ac} signal, and the \ac{smd} light source that was presented in Section~\ref{sec:smd_laser} to convert the electrical signal to an optical signal. The light from both \acp{ld} is coupled into an optical fibre for delivery to a launch optic. The receiver hardware is composed of a collimating lens to focus the detected light, a \ac{pin} detector, Femto receiver HSA-X-S-1G4-SI, to convert the optical signal to a photocurrent signal and a high-speed oscilloscope, Keysight MXR608A, to convert analogue waveforms back to digital signals.

Various communication signal processing, such as modulation, channel estimation and channel equalisation, are conducted on the software side by a \ac{pc}. The spectral efficient \ac{dco}-\ac{ofdm} with \ac{qam} modulations are implemented in this work \cite{9378787}. On the transmitter side, random information binary bits are generated and modulated as \ac{qam} symbols $X[k]$. To guarantee the time-domain signal to be real-valued, the frequency-domain \ac{qam} symbols need to fulfil the Hermitian symmetry, which requires $X[k]=X^{*}[K-k]$, where $k=0,1,\cdots,K-1$, $K$ is the length of the \ac{ofdm} symbol and $\{\cdot\}^{*}$ is the complex conjugate operator. This also implies that there are $K/2-1$ information carrying symbols in each \ac{ofdm} frame. Then, the frequency-domain signal can be converted to a time-domain signal via an \ac{ifft} operation \cite{6579696}:
\begin{align}
 x[n]=\frac{1}{\sqrt{K}}\sum_{k=0}^{K-1}X[k]\exp\left(\frac{2\pi jnk}{K}\right).
\end{align}
A \ac{cp} is added at the beginning of each \ac{ofdm} frame to deal with the non-flat channel and the interference between adjacent \ac{ofdm} frames. Pulse shaping has been also used to make the digital signal band-limited:
\begin{align}
x(t)=\sum_{n=0}^{K-1}{x}[n]p\left(t-nT_{\rm s}\right),
\end{align}
where $p(t)$ is defined as an \ac{rrc} signal pulse and $T_{\rm s}$ is the symbol period. The high amplitude samples of the time-domain signals are clipped to avoid excessive high \ac{papr} of the \ac{ofdm} signal:
\begin{align}
\hat{x}(t)=\left\{\begin{array}{lr} \sigma_{\rm x}\kappa &: {x}(t)\geq \sigma_{\rm x}\kappa \\
{x}(t) &: -\sigma_{\rm x}\kappa<{x}(t)<\sigma_{\rm x}\kappa \\
-\sigma_{\rm x}\kappa &:{x}(t)\leq -\sigma_{\rm x}\kappa\end{array}
\right.,
\label{eq:clipping1}
\end{align}
where $\sigma_{\rm x}$ is the standard deviation of ${x}(t)$ and $\kappa$ is the clipping level. After the signal clipping, the signal is forwarded to the \ac{awg} for data transmission.

At the receiver side, the received signal from the scope $y(t)$ is forwarded to a matched filter which downsample the signal to the symbol rate and removes the out-of-band noise:
\begin{align}
 y[n]=\{y \otimes p\}(t=nT_{\rm s}),
\end{align}
where $\otimes$ is the convolution operator. The \ac{smd} laser source has a very low input impedance (a few ohms) compared to the bias-T output impedance (50~$\Omega$). Consequently, it causes a severe impedance mismatch and signal reflections within the cable in-between. In addition, measurements showing significant nonlinearity exist in the optical link. In order to address these issues, cable length between bias-T and the \ac{smd} laser is minimised to avoid interference from the reflected signal in low frequency region. Furthermore, a Volterra nonlinear equaliser is implemented at the receiver side after the matched filter, as shown in Figure~\ref{fig:single_setup}. Therefore, the output of the nonlinear Volterra equaliser can be written as \cite{diniz1997adaptive}:
\begin{align}
z[n]&=\sum_{Q=1}^{Q_{\rm max}}\sum_{l_1=-L}^{L}\cdots\sum_{l_q=-L}^{L}\cdots\sum_{l_Q=-L}^{L}w_{l_1,\cdots,l_q,\cdots,l_Q}^{Q} \nonumber \\ &\times \prod_{q=1}^{Q} y[n-l_q],
\end{align}
where $Q$ is defined as the polynomial order and $L$ is half of the tap delay line, which corresponds to $2L+1$ taps in total. A maximum filter order of $Q_{\rm max}$ is considered. The conventional Volterra nonlinear equaliser has a very high complexity when $Q_{\rm max}$ is large. In many studies, simplified high-order nonlinear equalisers, such as memoryless polynomial, have been used to reduce the system complexity. In this work, we consider a modified Volterra nonlinear equaliser:
\begin{align}
z[n]&=\sum_{Q=1}^{Q_{\rm max}}\sum_{l_1=-L_Q}^{L_Q}\cdots\sum_{l_q=-L_Q,|l_q-l_1|\leq D_Q}^{L_Q}\cdots \nonumber \\ &\sum_{l_Q=-L_Q,|l_Q-l_1|\leq D_Q}^{L_Q}w_{l_1,\cdots,l_q,\cdots,l_Q}^{Q}\prod_{q=1}^{Q} y[n-l_q],
\label{eq:nleq_mod}
\end{align}
where terms of different orders depend on tap delay lines of different lengths. The tap delay line length of the $Q$th order is $L_Q$. The high order terms with products of samples with differences in delay greater than $D_Q$ are omitted. These changes can greatly reduce the number of high-order terms so that the computation complexity can be controlled. The modified nonlinear equaliser is trained by pilot symbols via a \ac{rls} algorithm defined in Algorithm~\ref{algo:RLS_algorithm}, where $\mathbf{y}_n$ is the nonlinear equalizer input polynomial vector:
\begin{algorithm}[t!]
	\SetAlgoLined
	\textbf{\textit{Initialisation:}} 
	$\mathbf{S}_{L_{\rm max}-1}=\frac{1}{\sigma_{x}^2}\mathbf{I}_{N_{\rm nl}}$, $\mathbf{w}_{L_{\rm max}-1}=\mathbf{0}_{N_{\rm nl}\times 1}$\\
	\For{$n=L_{\rm max},\cdots,L_{\rm max}+N_{\rm iter}$}{
		Updata tap delay line and $\mathbf{y}_n$ \\
		$e[n]=x[n]-\mathbf{y}_n^{\mathrm{T}}\mathbf{w}_{n-1}$ \\
		$\mathbf{s}_{n}=\mathbf{S}_{n-1}\mathbf{y}_n$ \\
		$\mathbf{S}_n=\frac{1}{\beta}\left(\mathbf{S}_{n-1}-\frac{\mathbf{s}_{n}\mathbf{s}_{n}^{\mathrm{T}}}{\beta+\mathbf{s}_{n}^{\mathrm{T}}\mathbf{y}_n}\right)$
		$\mathbf{w}_{n}=\mathbf{w}_{n-1}+e[n]\mathbf{S}_n\mathbf{s}_{n}$
	}
	\Return $\mathbf{w}_{L_{\rm max}+N_{\rm iter}}$
	\caption{Volterra equaliser RLS algorithm \cite{diniz1997adaptive}}
	\label{algo:RLS_algorithm}
\end{algorithm}
\begin{align}
&\mathbf{y}_n=\Big[y[n+L_1],\cdots,y[n-L_1],\cdots,y^2[n+L_2],\cdots,\nonumber \\
&\left. y[n-l_1]y[n-l_2],\cdots,y^2[n-L_2],\cdots,y^Q[n+L_Q],\cdots, \right.\nonumber \\
&\left. \prod_{q=1}^{Q} y[n-l_q],\cdots,y^Q[n-L_Q],\cdots,y^{Q_{\rm max}}[n+L_{Q_{\rm max}}],\cdots, \right.\nonumber \\
& \prod_{q=1}^{{Q_{\rm max}}} y[n-l_q],\cdots,y^{Q_{\rm max}}[n-L_{Q_{\rm max}}]
\Big]^{\mathrm{T}},
\end{align}
which include $N_{\rm nl}$ terms and $\mathbf{w}_n$ is the weight vector:
\begin{align}
&\mathbf{w}_n=\Big[w_{-L_1}^{1},\cdots,w_{L_1}^{1},\cdots,w_{-L_2,-L_2}^{2},\cdots,w_{l_1,l_2}^{2},\cdots, \nonumber \\
&\left. w_{L_2,L_2}^{2},\cdots,w_{-L_{Q},\cdots,-L_{Q}}^{Q},\cdots,w_{l_1,\cdots,l_{Q}}^{Q},\cdots,w_{L_{Q},\cdots,L_{Q}}^{Q}, \right.\nonumber \\
&\left. \cdots,w_{-L_{Q_{\rm max}},\cdots,-L_{Q_{\rm max}}}^{Q_{\rm max}},\cdots,w_{l_1,\cdots,l_{Q_{\rm max}}}^{Q_{\rm max}},\cdots,  \right.\nonumber \\
&w_{L_{Q_{\rm max}},\cdots,L_{Q_{\rm max}}}^{Q_{\rm max}}
\Big]^{\mathrm{T}}.
\end{align}
The time-domain \ac{ofdm} signal after pulse shaping (Figure~\ref{fig:single_setup}) is used as the reference signal $x[n]$ in Algorithm~\ref{algo:RLS_algorithm}. In this work, the nonlinear equaliser highest filter order $Q_{\rm max}$, number of taps $L_{Q}$ and the maximum delay difference $D_{Q}$ are optimised to achieve the highest data rate. Nonlinear equalisers of up to 5th order have been evaluated. In addition, the number of terms with different filter order is not greater than 50 so that the equaliser complexity is controlled in a feasible region. The redundant \ac{cp} is removed and followed by an \ac{fft} operation, which convert the time-domain signal to frequency-domain:
\begin{align}
Y[k]=\frac{1}{\sqrt{K}}\sum_{n=0}^{K-1}z[n]\exp\left(\frac{-2\pi jnk}{K}\right).
\end{align}
A one-tap equalisation is executed to retrieve the transmitted \ac{qam} symbols:
\begin{align}
 \tilde{X}[k] = Y[k]/{H}[k],
\end{align}
where ${H}[k]$ is the channel transfer function which can be estimated by sending pilot 4-\ac{qam} symbols with unit variance. Finally, the equalised symbols $\tilde{X}[k]$ are decoded and compared with the original binary bits for \ac{ber} calculation. 

In the modulation and demodulation process, the variance of the \ac{qam} symbols and the constellation order are adaptively selected according to the channel estimation information so that the available modulation bandwidth can be fully utilized and the achievable data rate can be maximised. In this work, the \ac{hh}-based adaptive bit and energy loading algorithm has been adopted, which has been widely used in many multi-carrier transmission systems \cite{hughes1987ensemble}.

\begin{table}[t]
	\caption{system setting for 905~nm \ac{ld}}
	\centering
	\begin{tabular}{c c}
		\hline
		\hline
		Parameters & Values\\
		\hline
		\ac{awg} peak-to-peak voltage & $140$~mV \\
		\ac{dc}-bias current  & 1171~mA \\
		Modulation bandwidth & 2.67~GHz \\
		Pulse shaping roll-off factor & 0.1 \\
		\ac{fft} size & 1024 \\
		\ac{cp} length & 20 \\
		Clipping level & 3.2 \\
		Highest nonlinear equaliser order & 5 \\
		Nonlinear equaliser number of taps for  \{2nd to 5th\} & $\{16,7,3,2\}$ \\
		Nonlinear equaliser largest delay difference \{2nd to 5th\} & $\{1,0,0,0\}$ \\
		\hline
		\hline
	\end{tabular}
	\label{table:Parameters_single_link}
\end{table}
In this work, \ac{smd} laser sources with ten different wavelengths for the communication \ac{ld} have been used. The communication channels corresponding to each wavelength shows a different characteristic. In this section, the performance results of the 905~nm \ac{ld} is presented. The used system settings are listed in Table~\ref{table:Parameters_single_link}, where the \ac{awg} peak-to-peak voltage and the \ac{dc}-bias current are carefully selected so that the average \ac{snr} across the modulation bandwidth is maximised. Note that the selection of modulation bandwidth can be greater than the channel bandwidth as the adaptive bit and energy loading function can avoid to load transmission resource to those subcarriers with very low \ac{snr}.

\begin{figure}[!t]
	\begin{center}
		\includegraphics[width=0.5\textwidth]{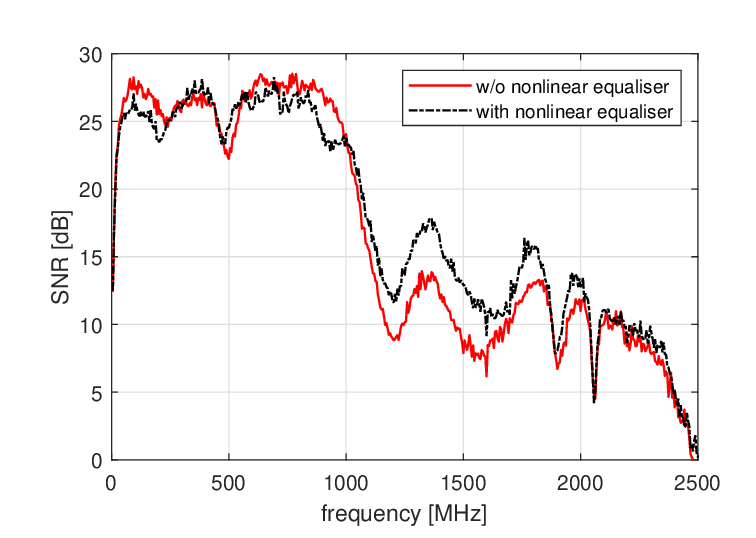}
	\end{center}
	\caption{Estimated \ac{snr} of \ac{lifi} link with 905~nm \ac{ld} on subcarriers of various frequencies with and without nonlinear equaliser.}
	\label{fig:single_SNR}
\end{figure}
\begin{figure}[!t]
	\begin{center}
		\includegraphics[width=0.5\textwidth]{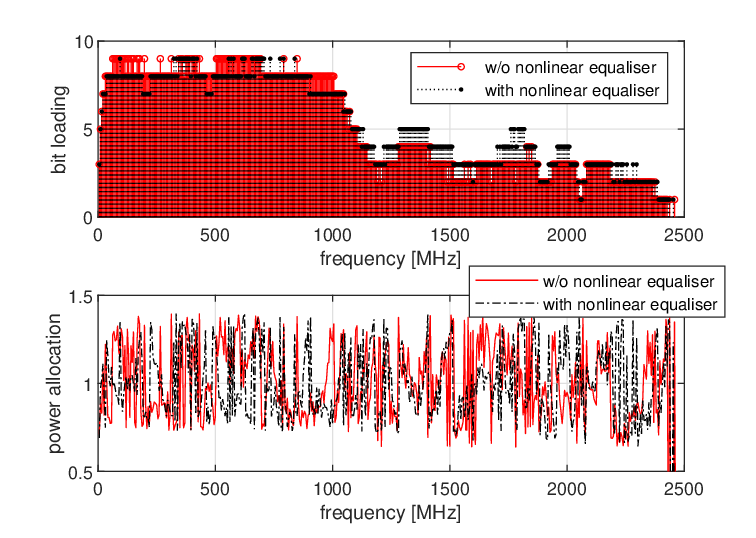}
	\end{center}
	\caption{An example of the adaptive bit (top figure) and energy (bottom figure) loading results.}
	\label{fig:single_loading}
\end{figure}
\begin{figure}[!t]
	\begin{center}
		\includegraphics[width=0.45\textwidth]{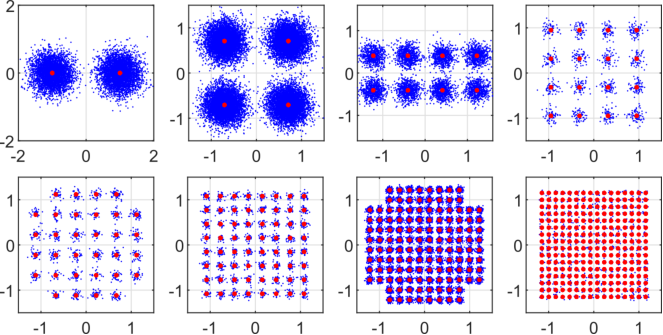}
	\end{center}
	\caption{Transmitted (red dots) and received (blue dots) \ac{qam} symbols of different modulation orders ($M=2,4,8,16,32,64,128,256,512$).}
	\label{fig:single_constellation}
\end{figure}
\begin{figure}[!t]
	\begin{center}
		\includegraphics[width=0.5\textwidth]{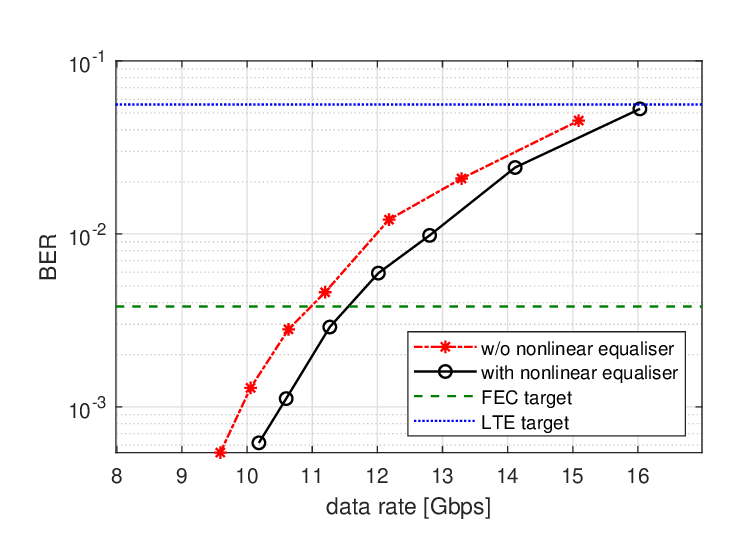}
	\end{center}
	\caption{\ac{ber} results against achievable data rate of \ac{lifi} link with 905~nm \ac{ld}.}
	\label{fig:single_BER_data_rate}
\end{figure}
\begin{figure*}[!t]
	\begin{center}
		\includegraphics[width=1\textwidth]{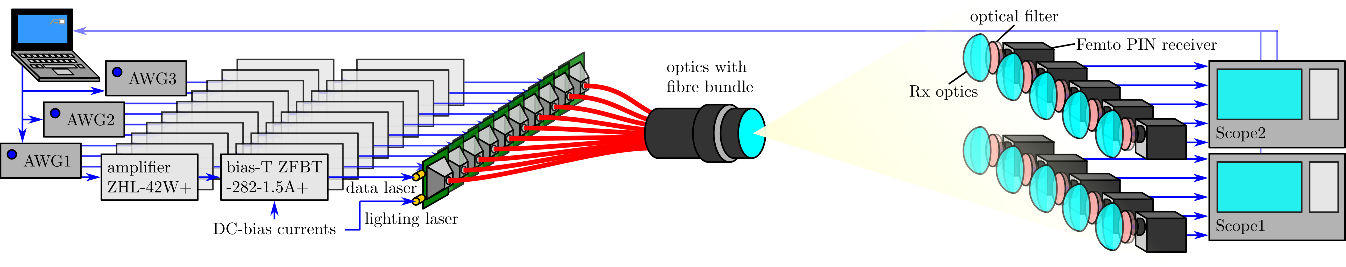}
	\end{center}
	\caption{Block diagram of \ac{lifi} transmission system with multiple parallel channels.}
	\label{fig:WDM_setup}
\end{figure*}

The results of \ac{snr} at various frequencies with a 905~nm \ac{smd} laser are shown in Figure~\ref{fig:single_SNR}. It can be observed that the achievable \acp{snr} below 1~GHz are above 20~dB, but \acp{snr} at a higher frequency drops significantly. This is due to a number of factors: 1. The used \ac{pin} receiver has a 3-dB bandwidth of 1.4~GHz and the responses beyond this frequency are severely degraded; 2. The system nonlinearity limits the achievable \acp{snr} at a high frequency; 3. The \ac{snr} is affected by the interference caused by signal reflections due to impedance mismatch. Despite the severe performance degradation at a high frequency, the maximum usable bandwidth is up to 2.5~GHz. It can also be observed that the link with nonlinear equaliser achieves higher \acp{snr} on some subcarriers, especially for high frequency subcarriers. However, it generally worsen the \ac{snr} at  very low frequency region. Overall, the nonlinear equaliser brings a boost to the high speed optical link quality. The high data rate transmission with up to 3~m was achieved with a single channel.

Based on the estimated channel and \acp{snr} shown in Figure~\ref{fig:single_SNR}, the adaptive bit and energy loading algorithm allocate different amounts of energy and numbers of bits to each subcarrier. One of the loading results is demonstrated in Figure~\ref{fig:single_loading}. It can be observed that the number of bits is highly related to the estimated \ac{snr}. The energy loading adjust the \ac{snr} to match the level that is just sufficient to transmit the allocated bit at the targeted \ac{ber}. The transmitted and received \ac{qam} symbols of different modulation orders are demonstrated in Figure~\ref{fig:single_constellation}, which demonstrated the successful equalisations of the symbols.

The results of resultant \ac{ber} against various achieved data rate are shown in Figure~\ref{fig:single_BER_data_rate}. It can be observed that the achievable data rate by the system with a nonlinear equaliser is slightly higher than that without a nonlinear equaliser, which is consistent with the \ac{snr} results shown in Figure~\ref{fig:single_SNR}. In this study, a \ac{ber} threshold of $5.6\times 10^{-2}$ is considered. This \ac{ber} threshold has been used in \ac{lte}, and it has been shown that with soft decision decoding and 3\% to 5\% overhead, such a \ac{ber} target is acceptable to reduce the final \ac{ber} below $1\times 10^{-6}$ \cite{6074908}. By considering such a \ac{ber} target, the achievable date rate can reach regions above 14~Gbps with 905~nm \ac{smd} laser source.


%
%

\begin{figure}[!t]
	\begin{center}
		\includegraphics[width=0.5\textwidth]{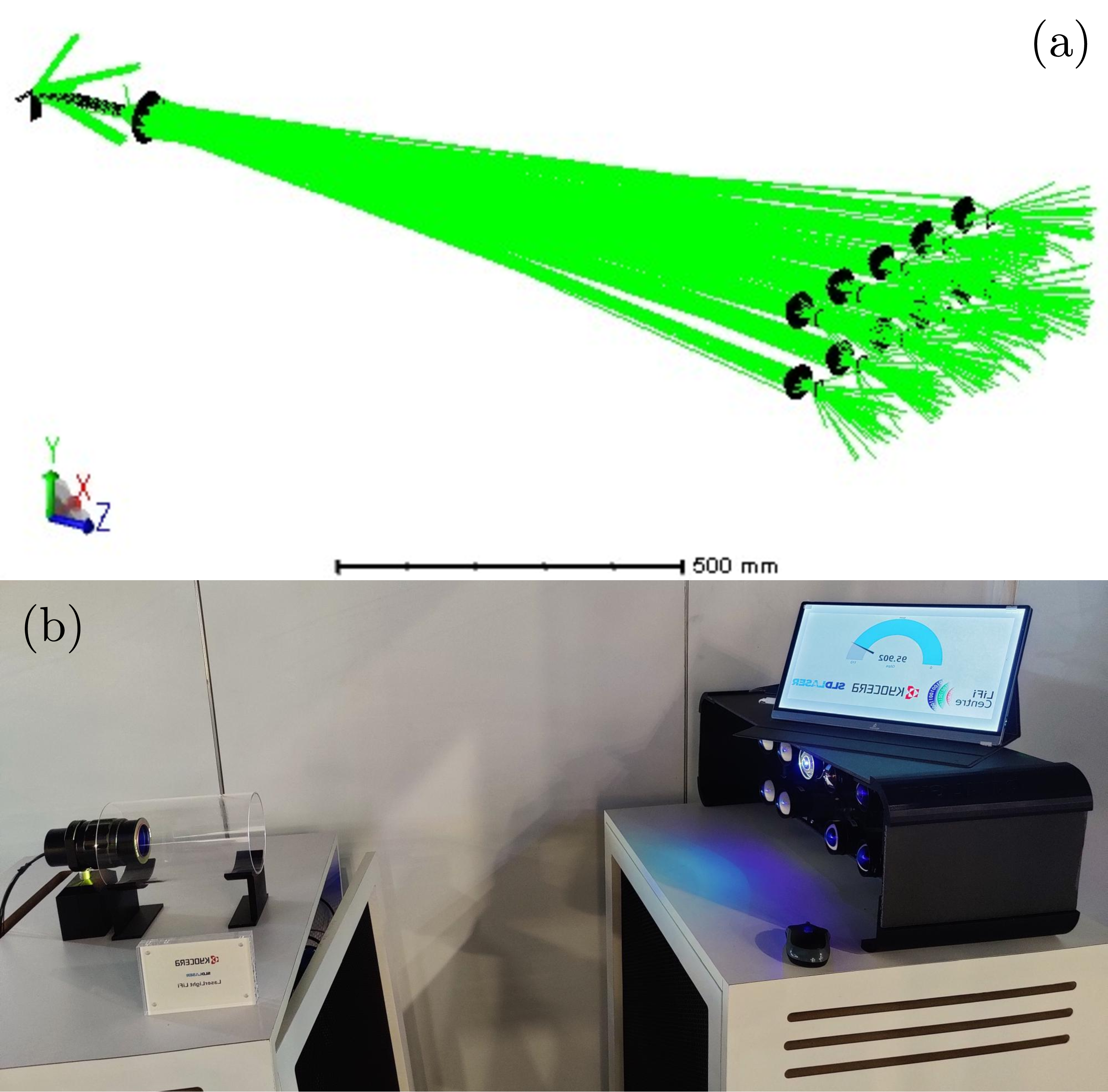}
	\end{center}
	\caption{Optical setup of the \ac{lifi} transmission system with ten parallel channels. (a) Zemax Opticstudio (b) Actual implementation}
	\label{fig:wdm_photo}
\end{figure}

\section{Over 100 Gbps LiFi transmission with parallel channels}
\label{sec:wdm_LiFi}
In Section~\ref{sec:single_LiFi}, the details about the \ac{lifi} system design for a single channel are introduced. On top of the single link design, a \ac{lifi} transmission system with ten parallel channels are presented in this section. The block diagram of the multi-channel \ac{lifi} transmission system is shown in Figure~\ref{fig:WDM_setup}. With regards to hardware part, ten copies of amplifiers and bias-Ts in conjunction with three 4-channel \acp{awg} are deployed to drive the \ac{smd} laser sources of different wavelengths. The output of the ten \ac{smd} laser sources are injected into a fibre bundle, which combines the optical signals and guides the light to a single launch optic. On the receiver side, an $2\times 5$ array of \ac{pin} \ac{pd} receivers are deployed. An optical filter with a 10~nm passband corresponds to each wavelength is mounted on each receiver to remove the crosstalk from other channels. Two 8-channel oscilloscopes are used to convert the photocurrent signals from ten receivers to digital signals. Therefore, the implemented system achieves parallel transmission via a \ac{wdm} approach. Regarding the wavelengths of the \acp{ld}, there are seven \ac{ir} devices: 850~nm, 900~nm, 905~nm, 940~nm, 955~nm, 980~nm, 1064~nm and three blue devices. For the case of transmission via \ac{ir} devices, the blue \ac{ld} mounted on the same device should also be activated by a \ac{dc} signal to provide white light illumination. For the case of transmission via blue devices, this is unnecessary as the communication \ac{ld} provides white light illumination, as well. To ensure the considered optical setup is feasible, a simulation in Zemax Opticstudio has been conducted to estimate the received optical power after each receiver lens. The optical setup defined in the simulation is illustrated in Figure~\ref{fig:wdm_photo}~(a). The simulation results show that the detected optical power after the receivers lens are in the range of a few to tens of mW, which is sufficient for reliable communication. An actual implementation of the optical system hardware is shown in Figure~\ref{fig:wdm_photo}~(b).

\begin{table*}[t]
	\caption{System setting and results for \ac{lifi} transmission with parallel channels}
	\centering
	\begin{tabular}{c c c c c c c c c c c c}
		\hline
		\hline
		Wavelengths [nm] &  405 & 450 & 455 & 850 &  900 & 905 & 940 & 955 & 980 & 1064 & Aggregate \\ 
		\hline
		\ac{awg} peak-to-peak voltage [mV] & 280 & 450 & 320 & 200 & 100 & 140 & 140 & $140$& 130 & 175 & ~\\
		\ac{dc}-bias current [mA]  & 930 & 1000 & 1050 & 1000 & 1100 & 1171& 1450 & 1450 & 1300 & 1050 & ~ \\
		Modulation bandwidth [GHz] & 1.33 & 1.6 & 1.6 & 2.67& 3 & 2.67& 2.67& 2.67& 2.67& 3 & ~ \\
		Clipping level & 3.4 & 3.2& 3.3& 3.2& 3.9 & 3.2& 3.2& 3.2& 3.2& 3.4 & ~ \\
		Highest nonlinear equaliser order & 5 & 0 & 3 & 0 & 3 & 5& 0 & 0 & 0 & 3 & ~ \\ 
		Data rate [Gbps] & 4.62 &  7.44 & 6.97 & 9.65 & 14.2 & 14.48 & 11.4 & 12 & 11.4 & 13.2 & 105.36 \\
		BER & 0.015 & 0.016 & 0.014 & 0.016& 0.015 & 0.028 & 0.009 & 0.007 & 0.007 & 0.018 & 0.0148 \\
		\hline
		\hline
	\end{tabular}
	\label{table:Parameters_WDM_link}
\end{table*}

With regards to software, the communication signal processing for each channel is identical to that introduced in Section~\ref{sec:single_LiFi} but with different setting parameters. In addition, the communication signal processing for all channels are executed in parallel. Due to the a number of undesired factors, such as temperature, nonlinearity, channel estimation error, the achieved \ac{ber} of each channel may deviate from the target level. Therefore, a simple adaptive algorithm is deployed to slightly adjust the bit loading solution so that the effective \ac{ber} converges to the target after a few iterations. The setting parameters of different channels are listed in Table~\ref{table:Parameters_WDM_link}. Except the \ac{fft} size, pulse shaping roll-off factor and \ac{cp} length, the setting parameters are adjusted to the optimised values so that the resultant achievable data rates are maximised. It can be observed that the channels with blue \acp{ld} require generally a high peak-to-peak voltage to overcome the stronger background noise in the blue spectrum region. Compared to channels with blue \acp{ld}, channels with \ac{ir} \acp{ld} use a higher bias current and a wider modulation bandwidth. Note that nonlinear equalisers are not used on five out of the ten channels, as shown in the 5th row of Table~\ref{table:Parameters_WDM_link}. This is because the use of a nonlinear equaliser brings negligible \ac{snr} improvement on these channels. 


\begin{figure}[!t]
	\begin{center}
		\includegraphics[width=0.5\textwidth]{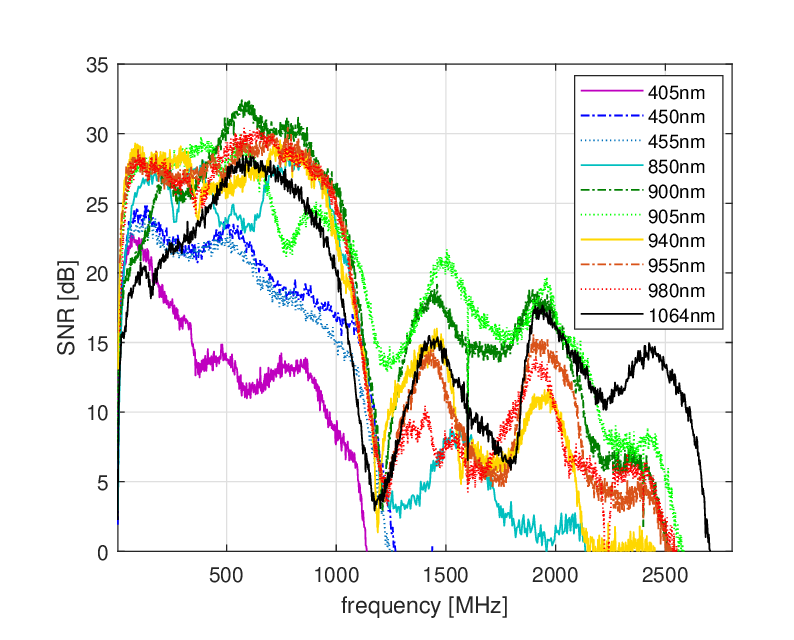}
	\end{center}
	\caption{\ac{snr} against frequency of \ac{lifi} transmission with multiple parallel channels.}
	\label{fig:WDM_SNR}
\end{figure}
\begin{figure}[!t]
	\begin{center}
		\includegraphics[width=0.5\textwidth]{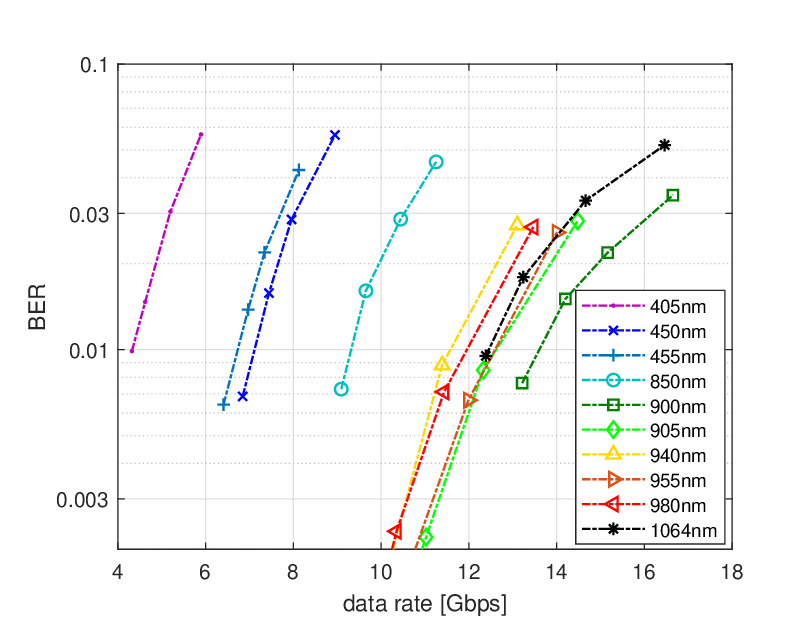}
	\end{center}
	\caption{\ac{ber} against achievable data rate of \ac{lifi} transmission with multiple parallel channels.}
	\label{fig:WDM_data_rate}
\end{figure}

The \ac{snr} and \ac{ber} results of all channels are shown in Figure~\ref{fig:WDM_SNR} and Figure~\ref{fig:WDM_data_rate}, respectively. It can be observed that the performance of different channels varies significantly due to the characteristics of each \ac{ld}. The three blue light channels achieve slightly worse link quality, where the usable bandwidths are above 1~GHz but lower than 1.5~GHz. The achievable \acp{snr} are in the range of 10 to 25~dB. Therefore, the achievable data rates are below 10~Gbps. In particular, the achievable data rate by 405~nm \ac{ld} is below 6~Gbps. In contrast, performance of links with \ac{ir} \acp{ld} are considerably superior, which generally exhibits a bandwidth of at least 2~GHz. In particular, the case with a 1064~nm laser uses a different \ac{pin} \ac{pd} receiver (HSA-X-S-2G-IN) with a wider 3-dB bandwidth of 2~GHz, which boosts the total usable bandwidth to about 2.7~GHz. In addition, the achievable \acp{snr} by \ac{ir} links at low frequency region (below 1~GHz) are in the range of 20 to 30~dB, while the achievable \acp{snr} at the high frequency region (above 1~GHz) are up to 20~dB, which lead to achievable data rates above 10~Gbps. The achievable data rates of the ten parallel channels are concluded in Table~\ref{table:Parameters_WDM_link}, which shows that the aggregated data rate of above 100~Gbps is achievable with a \ac{ber} lower than the \ac{lte} \ac{ber} target.

\begin{figure*}[!t]
	\begin{center}
		\includegraphics[width=1\textwidth]{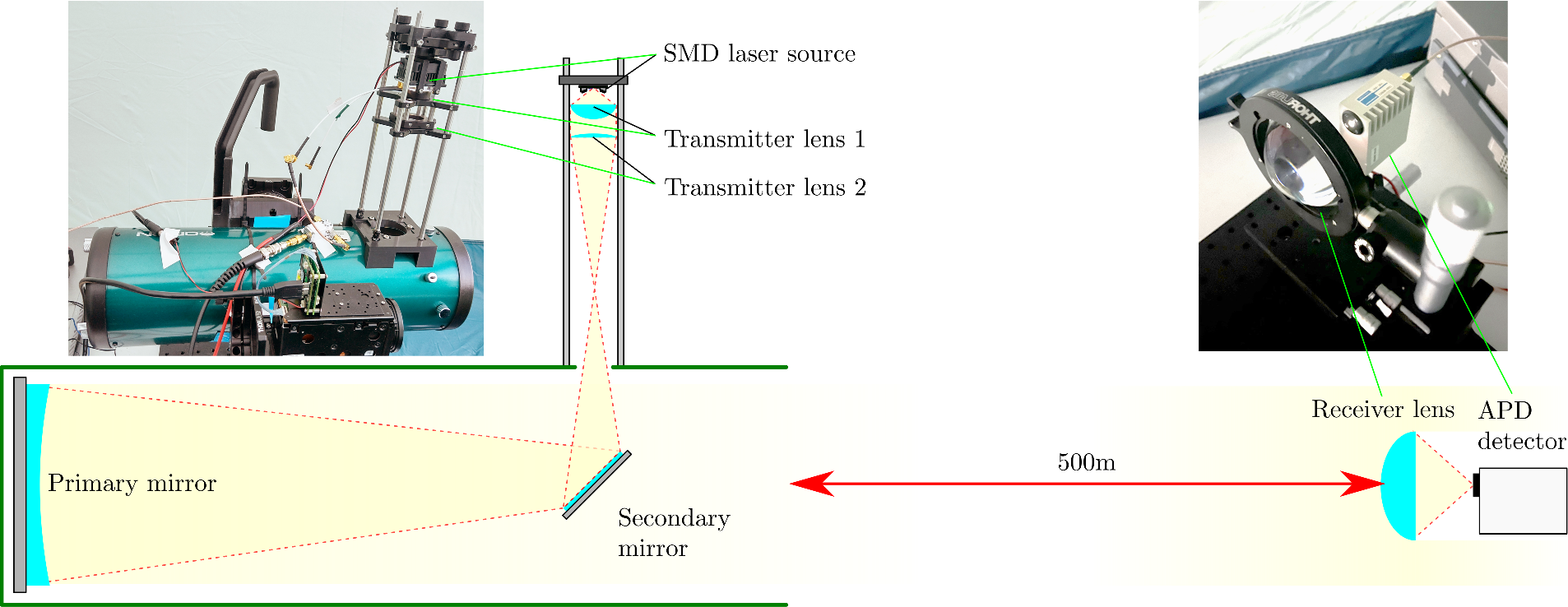}
	\end{center}
	\caption{Setup of the transmitter and the receiver for long-range \ac{lifi} link.}
	\label{fig:long_range_setup}
\end{figure*}
\begin{figure*}[!t]
	\begin{center}
		\includegraphics[width=1\textwidth]{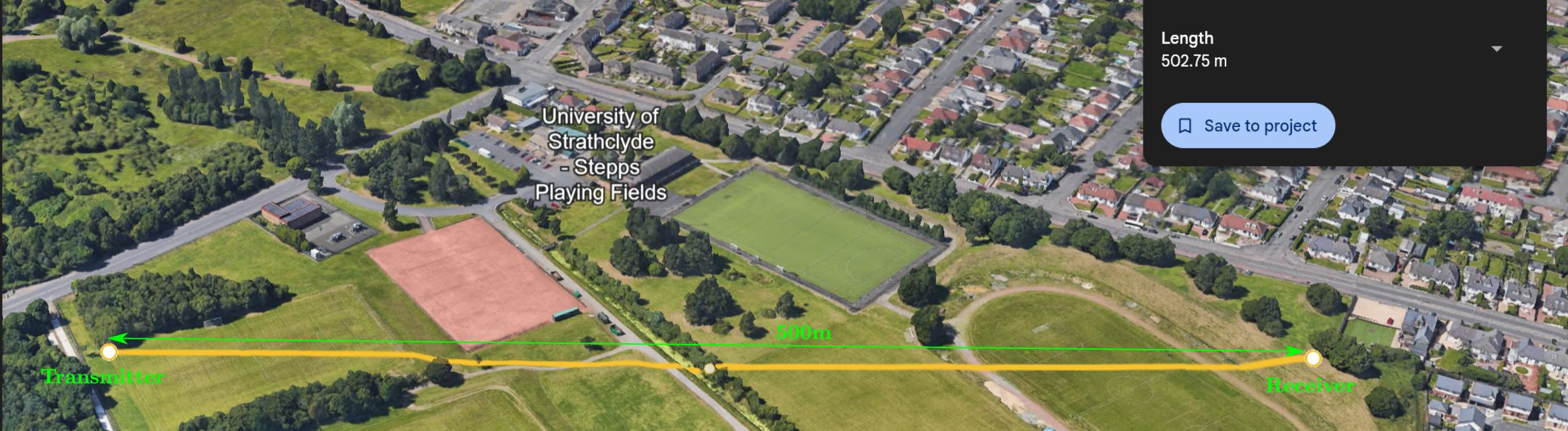}
	\end{center}
	\caption{Long range \ac{lifi} transmission measurement site on Google earth screenshot.}
	\label{fig:measurement_site}
\end{figure*}

\section{Long range LiFi transmission}
\label{sec:500m_LiFi}

As discussed in \ref{sec:use_cases}, the short range ultra high speed \ac{wdm} \ac{lifi} system will be useful in many indoor use cases. In addition, long range high speed \ac{lifi} transmission systems will also be attractive in vehicular communication and backhaul link use cases. Therefore, we showcase a long range \ac{lifi} transmission based on the \ac{smd} laser source over 500~m in this section. The optical setup of the \ac{lifi} system is illustrated in Figure~\ref{fig:long_range_setup}. On the transmitter side, a Newtonian telescope, Orion StarBlast 4.5 Astro Reflector Telescope, is used to launch the light. The telescope has a focal length of 450~mm, a focal ratio of $f/4$ and an aperture diameter of 113~mm. The \ac{smd} laser device is mounted on the top of the telescope eyepiece with a downward orientation. The \ac{smd} laser source emitting spot is positioned on the focal point of a 1-inch aspherical condenser lens with a \ac{na} of 0.79 (lens~1), which is used to capture most of the emitted light from the \ac{smd} laser device. A second plano-convex Lens (lens~2) is used to shape the light beam with the same focal ratio of the telescope. The focal point of the second lens and focal point of the telescope are collocated. Therefore, the output light of the telescope has a plane wavefront. Note that the secondary mirror of the telescope blocks some of the emitted light which is inevitable with a Newtonian telescope-based optical design. On the receiver side, a 75~mm diameter aspherical lens, Thorlab ACL7560U is used to capture the light after 500~m propagation distance. To compensate for the severe geometric loss over the long propagation distance, a high-gain \ac{apd} receiver, Hamamatsu C5658, is used to detect the captured light by the receiver lens. This \ac{apd} receiver has a 3-dB bandwidth of 1~GHz and an \ac{apd} gain of 100.

The long range measurement was conducted at the University of Strathclyde Stepps Playing Fields sport centre. The transmitter and the receiver are positioned at the two ends of the sport centre, as shown in the Google earth screenshot in Figure~\ref{fig:measurement_site}. It can be seen that the measured separation between the transmitter and the receiver on Google earth was a little more than 500~m. Note that the altitudes at different locations in the sport centre vary slightly. The transmitter and the receivers are located at the high grounds. Therefore, the trees and other objects at the low ground were not obstructing the \ac{los} path of the link.

\begin{figure*}[!t]
	\begin{center}
		\includegraphics[width=1\textwidth]{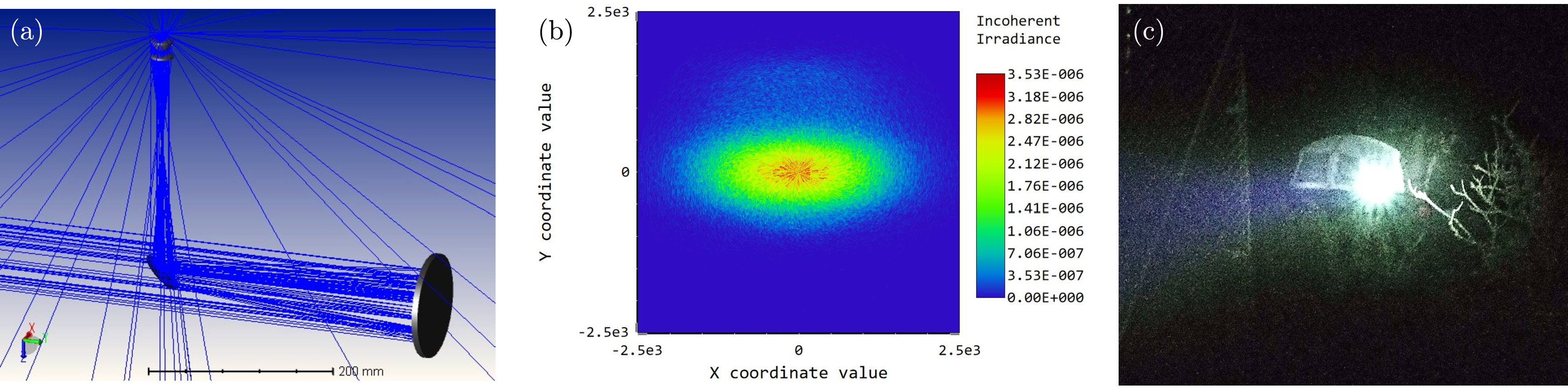}
	\end{center}
	\caption{(a) Zemax Opticstudio simulation setup (b) Incoherent irradiance light pattern at receiver plane obtained in simulation. (c) Actual light pattern at receiver plane taken by a camera.}
	\label{fig:light_pattern}
\end{figure*}
In order to justify that the considered optical setup is feasible to deliver long range wireless communication, a simulation in Zemax Opticstudio has been conducted. The optical setup is defined based on the specification shown in Figure~\ref{fig:long_range_setup} and a 5~m-by-5~m detector is defined at the receiver plane to capture the light pattern. The \ac{cad} model of the optical setup is illustrated in Figure~\ref{fig:light_pattern}~(a). The simulation result in terms of incoherent irradiance is shown in Figure~\ref{fig:light_pattern}~(b), where the light pattern at the receiver plane is about 3~m wide and about 2~m high. The actual light pattern has been captured by a camera during the night time, as shown in Figure~\ref{fig:light_pattern}~(c), which shows a close match to the simulation result.

\begin{table}[t]
	\caption{System setting and results for \ac{lifi} transmission over 500~m}
	\centering
	\begin{tabular}{c c c c}
		\hline
		\hline
		Wavelengths [nm] &  450 &  905 & Aggregate \\ 
		\hline
		\ac{awg} peak-to-peak voltage [mV] & 700 & 450 & ~ \\
		\ac{dc}-bias current [mA]  & 850 & 850 & ~ \\
		Modulation bandwidth [GHz] & 1.33 & 1.14 &  ~ \\
		Data rate [Gbps] & 2.41 & 2.43 & 4.84 \\
		BER & 0.0028 & 0.0035 & 0.0032 \\
		\hline
		\hline
	\end{tabular}
	\label{table:Parameters_500m_link}
\end{table}
The system parameter settings are summarised in Table~\ref{table:Parameters_500m_link}. In this experiment, a single \ac{smd} laser source is used, which is capable of emitting light in two wavelengths: 450~nm and 905~nm. Due to the significant geometric loss, the detectable optical power is much less than the short range scenario. Consequently, the performance of the optical links are limited by receiver noise. Therefore, higher peak-to-peak voltages of 700~mV and 450~mV are used for 450~nm and 905~nm channels, respectively. In addition, due to the restricted cooling methods in the outdoor scenario (a 5~cm-by-5~cm cooling fan), the bias current has to be limited to 850~mA. As shown in Figure~\ref{fig:single_SNR} and \ref{fig:WDM_SNR}, the links with the \ac{smd} laser sources exhibit higher \acp{snr} at low frequency regime ($<1$~GHz) and lower \acp{snr} at the high frequency regime ($>1$~GHz). Due to the lower received optical power, the overall signal quality is much worse than in the short-range scenario. Consequently, reliable transmission over high frequency subcarriers are not possible. Therefore, modulation bandwidth of 1.33~GHz and 1.14~GHz are used on 450~nm and 905~nm channels, respectively. Since the limiting factor is receiver noise, nonlinear equaliser is not used, which cannot effectively improve the performance.

\begin{figure}[!t]
	\begin{center}
		\includegraphics[width=0.49\textwidth]{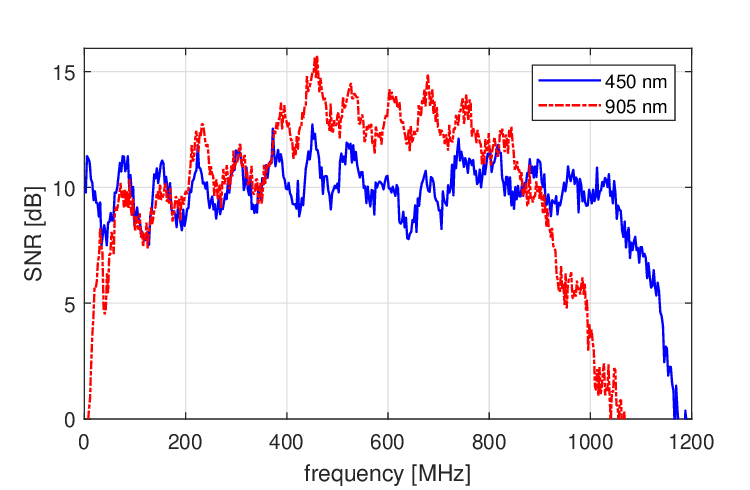}
	\end{center}
	\caption{Estimated SNR against frequency in the long range \ac{lifi} link.}
	\label{fig:long_range_SNR}
\end{figure}
\begin{figure}[!t]
	\begin{center}
		\includegraphics[width=0.49\textwidth]{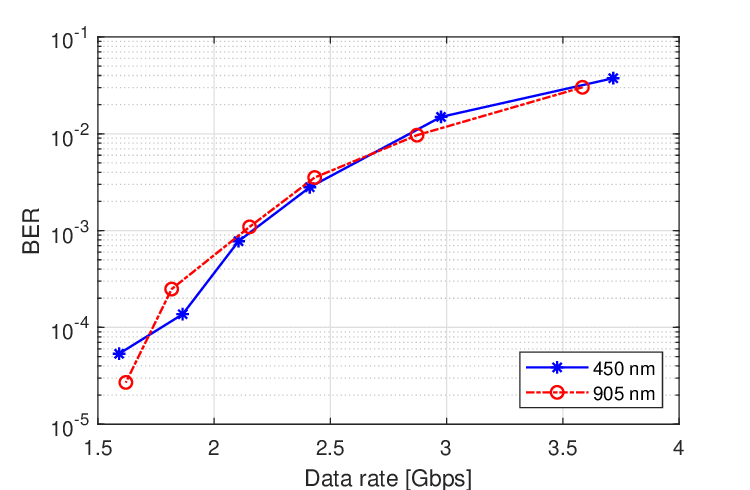}
	\end{center}
	\caption{BER against data rate in the long range \ac{lifi} link.}
	\label{fig:long_range_BER}
\end{figure}
The long range communication performance is concluded in Figure~\ref{fig:long_range_SNR} and Figure~\ref{fig:long_range_BER}. Compared to the short range scenario, the achievable \ac{snr} in the long-range case decreases to a range of 5 to 15~dB due to the decreased level of received optical power, as shown in Figure~\ref{fig:long_range_SNR}. In the case of 450~nm channel, the usable modulation bandwidth is up to 1.2~GHz. The estimated \ac{snr} is around 10~dB over the used modulation bandwidth. In the case of the 905~nm channel, a slightly narrower modulation bandwidth of 1.05~GHz has been achieved. However, for the subcarriers in the frequency range of 400~MHz to 800~MHz, the achievable \acp{snr} are higher than those on the 450~nm channel by 2 - 3~dB. In terms of \ac{ber} against achievable data rate, the performance of links with the two channels is similar, as shown in Figure~\ref{fig:long_range_BER}. The 450~nm channel can achieve a data rate of 2.41~Gbps with a \ac{ber} of 0.0028 and the 905~nm channel can achieve a data rate of 2.43~Gbps with a \ac{ber} of 0.0035. Therefore, the aggregate data rate is about 4.84~Gbps with an effective \ac{ber} of 0.0032, which is lower than the \ac{fec} target. This long-range \ac{lifi} link demonstrate that the \ac{smd} laser device can be used to achieve multi-Gbps data rate \ac{lifi} transmission in a long range distance of at least 500~m.

\section{Conclusions}
\label{sec:conclusion}
In this paper, we have discussed the potential use cases of laser-based \ac{lifi} systems. We have demonstrated a \ac{lifi} \ac{wdm} system based on ten \ac{smd} laser sources achieving over 100~Gbps data rate and a multi-Gbps long range \ac{lifi} transmission over a distance of 500~m. In particular, we have demonstrated the use of a nonlinear equaliser to improve transmission channel quality. The experimental transmission performance in terms of \ac{snr}, \ac{ber} and achievable data rates are presented. The demonstrated system proves that it is possible to scale the transmission capacity of the \ac{lifi} system by using multiple \ac{smd} laser sources of different wavelengths. In addition, it is also possible to utilise the high optical power feature of the \ac{smd} laser source to establish optical links over long distances. 

\bibliographystyle{IEEEtran}
\bibliography{KSLD_100G_ref.bib}

\end{document}